\documentclass[runningheads,letterpaper]{llncs}
\usepackage[T1]{fontenc}
\usepackage[utf8]{inputenc}
\usepackage{lmodern}
\usepackage{amsmath,amssymb,graphicx,booktabs,url,enumitem}
\usepackage{hyperref}
\setlist{leftmargin=13pt,topsep=6pt,itemsep=0pt,parsep=0pt,partopsep=0pt}
\hypersetup{hidelinks,pdftitle={Visible Reasoning: User-Facing Decision Transparency for Generative AI Systems},pdfauthor={Sanjay Krishna Anbalagan and coauthors}}
\newcommand{\origpage}{}
\newcommand{\FA}{\mathrm{FA}}
\newcommand{\AVG}{\mathrm{AVG}}
\newcommand{\PERSIST}{\mathrm{PERSIST}}
\newcommand{\DEPTH}{\mathrm{DEPTH}}
\newcommand{\simfn}{\operatorname{sim}}
\begin{document}
\title{Visible Reasoning: User-Facing Decision\\Transparency\\for Generative AI Systems}
\titlerunning{Visible Reasoning}
\author{Sanjay Krishna Anbalagan\inst{1} \and Xinrui Nie\inst{1} \and Anughna Kommalapati\inst{1} \and Vijay Kumar Kanamarlapudi\inst{1} \and Sreekanth Radhakrishnan\inst{1} \and Xiaodan Zhao\inst{1} \and Umesh Mohan\inst{1}}
\authorrunning{S.K. Anbalagan et al.}
\institute{Amazon Web Services, USA}
\maketitle
\begin{abstract}
Across multiple waves of computing, public trust in new technology has followed the emergence of governance mechanisms that made dominant risks legible and manageable. Generative AI shifts this pattern: these systems synthesize across heterogeneous sources, and users increasingly experience them as reasoning systems. The central question changes from “did it perform correctly?” to “can we trust how it arrived at its conclusion?”

This paper introduces \textbf{Filtered Decision Logs (FDLs)}, a user-facing transparency mechanism that makes decision formation visible without exposing raw chain-of-thought. FDLs pair each response with (i) a concise decision rationale, (ii) \textbf{influence-ordered supporting data points} derived from tool outputs, and (iii) relevance indicators showing which sources most shaped the conclusion. To operationalize evidence selection, we propose a trace-based method that reconstructs observability telemetry into a provenance tree and ranks evidence using four deterministic semantic and structural signals---\textbf{Final Answer Similarity, Average Relevancy, Persistence, and Structural Proximity}---combined via adaptive weighted scoring. We demonstrate FDLs across \textbf{three domains}: career advising (5 tools), investment analysis (4 tools), and travel planning (4 tools). In a formative evaluation with \textbf{50 participants} (25 students, 25 working professionals in AI development), \textbf{92\%} rated the influence-ranked evidence as helpful for understanding and trusting the system's recommendation. \textbf{Counterfactual tool-removal experiments} confirm that removing the top-ranked source changes the answer more than removing a low-ranked source, validating that the ranking tracks decision dependence. We conclude that evidence-ranked transparency helps users identify misleading signals and missing context, enabling more confident acceptance, correction, or re-run of system decisions.
\end{abstract}
\origpage
\section{Introduction}
New technology is rarely trusted simply because it exists; it becomes trusted when its risks can be governed. In earlier eras, governance co-evolved with adoption. Reliability concerns about mechanized work led to audits and verification; privacy concerns about networked information led to encryption and consent frameworks; and security concerns about cloud services led to compliance standards and operational controls. In each case, governance made risk tractable, and trust followed.

Generative AI changes the surface area of uncertainty. The user is not judging a deterministic procedure but a system that synthesizes across heterogeneous sources and produces persuasive outputs under incomplete information. For many tasks, users cannot easily validate correctness, and it is often ambiguous whether a failure arose from missing context, irrelevant evidence, tool errors, or flawed reasoning. Output-only interfaces force a brittle trust posture: either the user accepts an answer with insufficient grounds or rejects it without knowing how to correct it.

This challenge becomes sharper in modern agentic applications. Tool-augmented LLM agents~\cite{react,toolformer} call external tools---APIs, databases, search engines---to gather real-time data before producing a response. When an investment advisor agent synthesizes stock prices, financial news, company fundamentals, and economic indicators into a recommendation, the user faces a fundamental question: \emph{which data sources actually shaped this answer, and how much?} Without this information, the user cannot verify whether the recommendation is grounded in solid fundamentals or recent news hype, cannot challenge the recommendation by identifying potentially misleading sources, and cannot re-run the agent with different data source configurations to explore alternatives.

A human analogy clarifies the expected interaction. When people ask a parent for career advice, trust is grounded in shared context; when they ask an expert, trust is contingent on justification---users expect data points and reasoning before accepting the recommendation. In expert settings, transparency is not merely an explanation; it is a mechanism for verification, disagreement, and refinement. We argue that generative AI systems should support a comparable interaction: users should be able to inspect the evidence that shaped an answer and decide whether to accept it, provide missing context, or re-run the system with corrected constraints.

This paper introduces \emph{Filtered Decision Logs (FDLs)} as a user-facing transparency artifact that provides evidence-backed decision transparency without exposing raw chain-of-thought. FDLs are built for the practical reality of modern systems: many generative applications already use tools and sub-agents, and many platforms already provide observability telemetry~\cite{otel}. The missing piece is a method that transforms this telemetry into user-comprehensible, influence-ranked evidence---a “governance surface” that lets users evaluate the logic behind outputs rather than validating outputs alone.
\origpage
\subsection{Contributions}
This paper makes four contributions:
\begin{enumerate}
\item \textbf{Filtered Decision Logs}: a user-facing transparency artifact consisting of decision rationale, influence-ordered supporting data points from tool outputs, and relevance indicators (Section~\ref{sec:fdl}).
\item \textbf{Provenance reconstruction}: a trace-based method that reconstructs an agent run into a provenance tree spanning LLM steps and tool artifacts (Section~\ref{sec:architecture}).
\item \textbf{Deterministic influence ranking}: a scoring method combining four semantic and structural signals with adaptive weighting, validated by counterfactual tool-removal experiments (Section~\ref{sec:scoring}).
\item \textbf{Multi-domain prototype and formative evaluation}: a research prototype demonstrated across three domains with formative feedback from 50 participants (Section~\ref{sec:demonstration}--\ref{sec:evaluation}).
\end{enumerate}
\section{Related Work}
Our work intersects three research threads---training-data influence, explanation and rationale methods, and provenance and observability---but focuses on a distinct object: runtime evidence influence in tool-using agent traces, surfaced as a user-facing governance interface.
\subsection{Training-Data Influence and Data Valuation}
Influence functions, introduced by Koh and Liang~\cite{koh}, trace a model's prediction back to training examples most responsible for that prediction. Data Shapley, proposed by Ghorbani and Zou~\cite{shapley}, evaluates training-data value using Shapley-based allocation principles. These methods answer a different question than ours: in tool-using generative systems, users need to know which runtime artifacts---retrieved documents, API responses, database rows---shaped the answer in the current run, not which training examples shaped the model's parameters. Our influence scoring operates at the runtime tool-output level, complementing rather than replacing training-data attribution.
\subsection{Rationales, Interpretability, and Chain-of-Thought}
Rationale extraction methods, such as those proposed by Lei et al.~\cite{rationalizing}, learn to identify concise text fragments that justify a prediction. In large language models, chain-of-thought (CoT) prompting, introduced by Wei et al.~\cite{cot}, improves performance by eliciting intermediate reasoning steps. Wang et al.~\cite{selfconsistency} further improved CoT by sampling multiple reasoning paths and selecting the most consistent answer.

These approaches improve reasoning performance and can provide natural-language explanations, but they do not directly solve the governance question
\origpage
\noindent in tool-using systems: which external artifacts influenced the answer, and how strongly? Moreover, exposing raw chain-of-thought is not necessarily suitable as a user-facing governance mechanism. Turpin et al.~\cite{unfaithful} demonstrated that CoT traces do not always faithfully represent the model's actual decision process---models can produce coherent reasoning chains that lead to the correct answer for the wrong reasons. For governance purposes, self-narrated reasoning is a status update, not auditable evidence.

We draw on the concept of user-in-the-loop explanation interfaces. Chromik et al.~\cite{illusion} showed that users can overestimate their understanding from AI explanations alone, motivating more evidence-grounded transparency mechanisms. FDLs address this by providing quantified, verifiable evidence items rather than narrative explanations that may create an illusion of understanding.
\subsection{Feature Attribution and Explainability}
Feature attribution methods such as LIME, introduced by Ribeiro et al.~\cite{lime}, and SHAP, proposed by Lundberg and Lee~\cite{shap}, explain individual predictions by approximating local model behavior. LIME generates local linear approximations around a prediction to identify which input features contributed most, while SHAP uses Shapley values from cooperative game theory to provide theoretically grounded feature importance scores. These methods operate at the token or feature level; FDLs operate at the \emph{tool level}, making them interpretable by non-technical users who understand “the news article had a 0.64 influence score” but not “token 47 had attention weight 0.03.”
\subsection{Provenance, Observability, and Evaluation}
Distributed tracing standards, such as OpenTelemetry~\cite{otel}, model execution as connected parent--child spans within a trace. Emerging frameworks like Langfuse~\cite{langfuse} emphasize continuous, introspectable trace capture by instrumenting agents at runtime. Evaluation frameworks such as RAGAS, proposed by Es et al.~\cite{ragas}, provide automated evaluation of retrieval-augmented generation using LLM-as-evaluator patterns. Our work complements these efforts by transforming connected traces into influence-ordered evidence per answer---a specific transparency artifact designed for end users rather than developers.
\subsection{Why Not Ask the Model?}
A natural question is: why not simply ask the LLM which evidence it used? We distinguish three paradigms for decision transparency based on \emph{evaluation independence}---whether the transparency mechanism depends on the same generative process it is meant to make transparent (Table~\ref{tab:paradigms}).

The analogy is a financial audit: the auditor does not ask the company “were your transactions legitimate?”---they examine the receipts, trace the money flow, and independently verify what happened. FDLs provide this evaluation independence: the scoring mechanism is entirely separate from the reasoning mechanism.
\origpage
\begin{table}[!ht]
\caption{Three paradigms for AI decision transparency. Evaluation independence determines suitability for governance.}\label{tab:paradigms}
\centering
\begin{tabular}{lccc}
\toprule
Paradigm & Deterministic & Verifiable & Governance\\
\midrule
LLM-as-evaluator~\cite{ragas} & No & No & Low\\
Chain-of-thought~\cite{cot} & No & No & Low\\
FDLs (external measurement) & Yes & Yes & High\\
\bottomrule
\end{tabular}
\end{table}
A user can trust the influence score even if they do not trust the recommendation.
\section{Filtered Decision Logs as a Governance Interface}\label{sec:fdl}
We define a \emph{Filtered Decision Log (FDL)} as a transparency artifact displayed alongside a model response. An FDL answers three user questions: (1) What is the system's conclusion? (2) Why did it reach that conclusion? (3) What evidence most shaped the conclusion?

\begin{figure}[!ht]
\centering\includegraphics[width=0.65\textwidth]{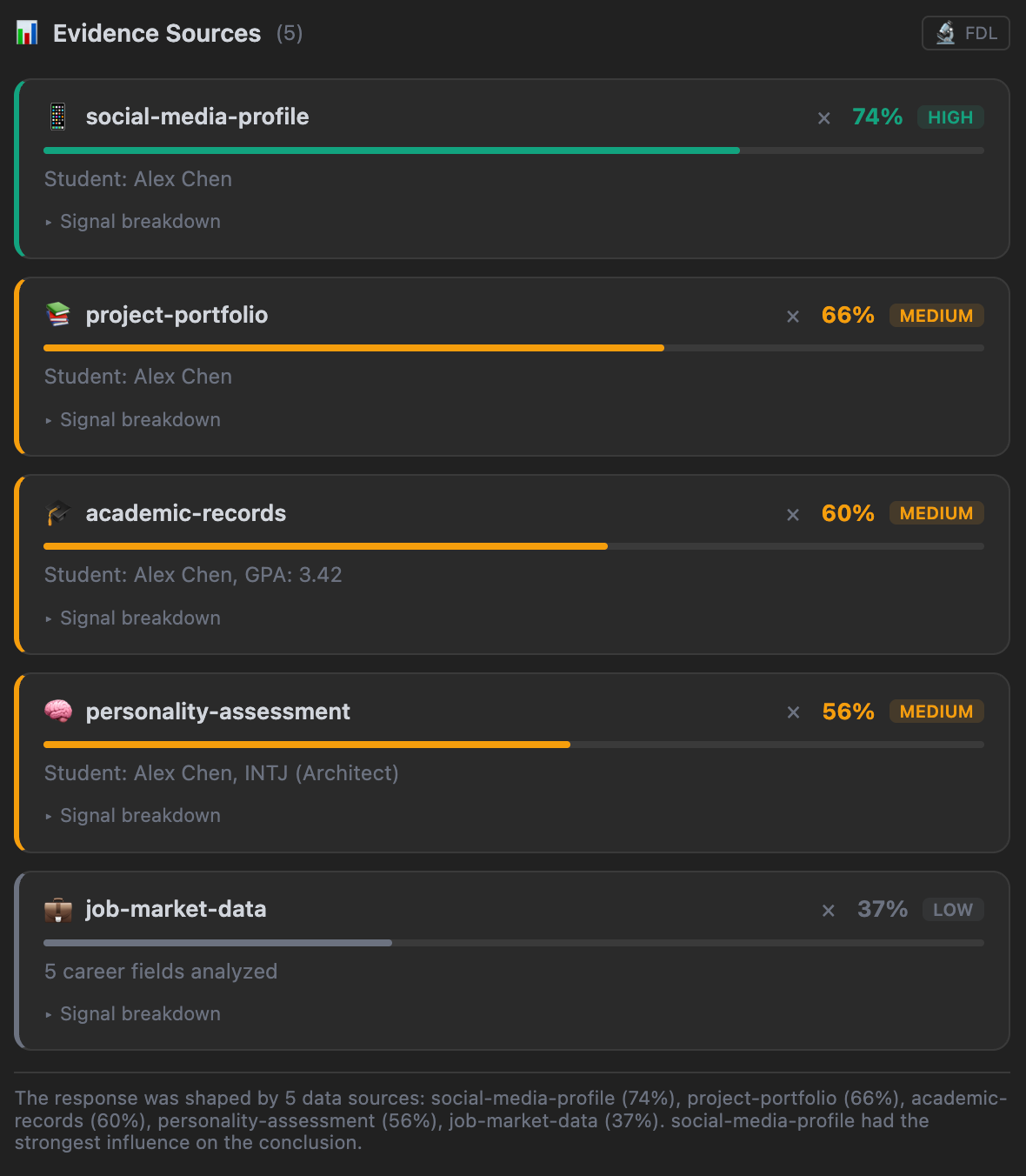}
\caption{FDL influence panel for the career advisor scenario. Five evidence sources ranked by influence score: social-media-profile 0.74 (High), project-portfolio 0.66, academic-records 0.60, personality-assessment 0.56 (all Medium), and job-market-data 0.37 (Low). Users can expand signal breakdowns or exclude sources for re-asking.}\label{fig:career}
\end{figure}

FDLs operationalize this through three components:
\paragraph{Decision rationale.} A concise natural-language summary describing why the answer was produced, anchored in evidence items rather than raw intermediate reasoning.
\paragraph{Supporting data points.} Derived from tool outputs---retrieved documents, API responses, database results. Instead of exposing raw JSON traces, the system presents readable evidence statements with semantic descriptions (e.g., “NVIDIA Corporation: Revenue \$130.5B, Margin 55.9\%”).
\paragraph{Relevance indicators.} Categorical labels---High ($\geq 0.70$), Medium (0.40--0.70), Low ($<0.40$)---grounded in the deterministic scoring method described in Section~\ref{sec:scoring}. Color coding provides immediate visual triage.
\paragraph{Interaction pattern: Accept or Re-Run.} FDLs support an accept/re-run interaction (Fig.~\ref{fig:career}). The user inspects the influence ranking, optionally expands signal breakdowns, and if a source appears misleading, excludes it and re-runs the agent. The system generates a new recommendation with fresh influence scores. This transforms governance from a hidden internal process into a collaborative loop.
\paragraph{Important clarification on scores.} Influence scores are absolute alignment and proximity measures computed independently for each evidence item. They do not sum to 1.0 or 100\%, and they are \emph{not causal attribution}---a score of 0.70 means “high semantic alignment with the answer and reasoning chain,” not “this source contributed 70\% of the answer.” This distinction is critical: the scores indicate how much a source's content appears in the output, not a causal decomposition of the generative process.
\origpage
\section{System Architecture}\label{sec:architecture}
The FDL framework consists of two independent pipelines. This separation is the foundation of evaluation independence: the agent never knows it will be evaluated, and the FDL pipeline never influences the agent's reasoning.
\subsection{Agent Pipeline}
The agent pipeline receives a user query, calls external tools, reasons over the results through multiple iterations, and produces a final answer. The execution trace is recorded as a \emph{context tree}---a hierarchical data structure where each node represents a pipeline stage. Our prototype uses a tool-augmented LLM agent built on Amazon Bedrock (Claude 3 Sonnet) with the ReAct pattern~\cite{react}: the agent iteratively reasons about which tools to call, executes them, observes results, and continues until it produces a final answer.
\origpage
\begin{figure}[!ht]
\centering\includegraphics[width=0.22\textwidth]{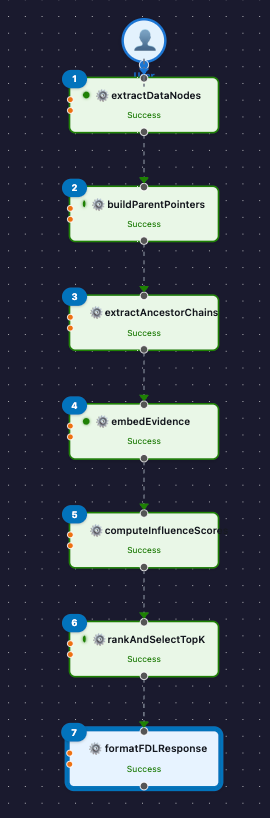}
\caption{The 7-stage FDL influence pipeline. Each stage is a pure function operating on a typed scope. The pipeline uses no generative AI---only embeddings (Stage 4) and deterministic computation.}\label{fig:pipeline}
\end{figure}
\subsection{FDL Pipeline}
The FDL pipeline is a separate 7-stage linear pipeline (Fig.~\ref{fig:pipeline}) that processes the context tree using \emph{no generative AI}---only embeddings and deterministic computation:
\begin{enumerate}
\item \textbf{extractDataNodes}: Walk the execution trace to identify tool result stages.
\item \textbf{buildParentPointers}: Construct a parent pointer map for upward traversal.
\item \textbf{extractAncestorChains}: For each data node, collect LLM reasoning ancestors with text and depth.
\item \textbf{embedEvidence}: Compute embedding vectors using Amazon Titan Embed Text v2~\cite{titan}. A cache prevents redundant API calls.
\item \textbf{computeInfluenceScores}: Calculate four signals and combine into composite scores (Section~\ref{sec:scoring}).
\item \textbf{rankAndSelectTopK}: Sort by score descending, select top-$K$, assign indicators.
\item \textbf{formatFDLResponse}: Produce the final FDL with rationale, ranked evidence, and signal breakdowns.
\end{enumerate}
\subsection{Provenance Reconstruction}
The pipeline begins by reconstructing a provenance tree from the agent's execution trace. Modern agent frameworks record execution as connected parent--child
\origpage
\noindent spans~\cite{otel,langfuse}. We walk this tree recursively to identify data nodes---stages whose logs contain tool result data.

A practical challenge is that tool execution often occurs dynamically within a single orchestration stage. We address this with a fallback extraction strategy: when the tree walk finds fewer data nodes than expected, the system extracts tool results directly from the global conversation messages array. This ensures framework-agnostic provenance reconstruction regardless of how the agent structures its execution trace.
\section{Deterministic Influence Scoring}\label{sec:scoring}
Each evidence item is scored on four orthogonal signals computed using cosine similarity between embedding vectors produced by Amazon Titan Embed Text v2~\cite{titan}, ensuring determinism---the same inputs always produce the same scores. Unlike LLM-as-evaluator approaches~\cite{ragas} where re-running the same evaluation can produce different results, FDL influence scores are reproducible.
\subsection{Signal Definitions}
Let $e_d$ denote the embedding of data node $d$, $e_f$ the embedding of the final answer, and $\{e_{a_1},\ldots,e_{a_n}\}$ the embeddings of $d$'s LLM ancestor stages. Let $\simfn(u,v)=\frac{u\cdot v}{\lVert u\rVert\,\lVert v\rVert}$ denote cosine similarity.
\paragraph{Final Answer Similarity (FA).} Direct semantic overlap between the tool result and the final answer:
\begin{equation}\label{eq:fa}
\FA(d)=\simfn(e_d,e_f)
\end{equation}
This is the strongest signal: if a tool's output appears verbatim or paraphrased in the answer, FA will be high. Conceptually related to faithfulness metrics in RAG evaluation~\cite{ragas}, but computed deterministically via embeddings rather than LLM judgment.
\paragraph{Average Relevancy (AVG).} Mean semantic similarity between the tool result and each LLM reasoning step:
\begin{equation}\label{eq:avg}
\AVG(d)=\begin{cases}
\frac{1}{n}\sum_{i=1}^{n}\simfn(e_d,e_{a_i}) & \text{if }n>0\\
0 & \text{if }n=0
\end{cases}
\end{equation}
Captures whether the data was consistently relevant across the agent's reasoning chain.
\paragraph{Persistence (PERSIST).} Fraction of ancestor similarities exceeding a threshold $T$:
\begin{equation}\label{eq:persist}
\PERSIST(d)=\begin{cases}
\dfrac{\left|\{a_i:\simfn(e_d,e_{a_i})>T\}\right|}{n} & \text{if }n>0\\
0 & \text{if }n=0
\end{cases}
\end{equation}
\origpage
\noindent Unlike AVG which averages, PERSIST measures \emph{breadth} of influence. A tool referenced in 4 of 5 reasoning steps ($\PERSIST=0.8$) scores higher than one referenced intensely in only 1 step. Default $T=0.30$. Inspired by the persistence concept in influence functions~\cite{koh}, adapted from training-time to runtime evidence.
\paragraph{Structural Proximity (DEPTH).} Inverse of the number of LLM reasoning ancestors:
\begin{equation}\label{eq:depth}
\DEPTH(d)=\frac{1}{1+n}
\end{equation}
A direct tool result with no intermediaries gets $\DEPTH=1.0$. We count only LLM reasoning ancestors (stages with text output), not pipeline plumbing stages, to avoid diluting the signal with framework overhead.
\subsection{Composite Score}
The four signals are combined with configurable weights:
\begin{equation}\label{eq:score}
S(d)=\alpha\cdot\FA(d)+\beta\cdot\AVG(d)+\gamma\cdot\PERSIST(d)+\delta\cdot\DEPTH(d)
\end{equation}
Default weights: $\alpha=0.40$, $\beta=0.30$, $\gamma=0.20$, $\delta=0.10$, summing to 1.0. The weights encode a prior: direct semantic similarity (FA) is the strongest signal, followed by consistency across reasoning (AVG), breadth of reference (PERSIST), and structural proximity (DEPTH). This weighted combination is conceptually related to how SHAP~\cite{shap} combines feature contributions, but operates at the tool level with configurable weights rather than Shapley-value-based allocation.
\paragraph{Scores are not additive.} Each evidence item's score is computed independently. Scores do not sum to 1.0 across items and do not represent causal attribution. A score of 0.70 means “high semantic alignment with the answer,” not “this source contributed 70\% of the answer.”
\subsection{Adaptive Weight Redistribution}
The default weights assume a multi-step reasoning pipeline where all four signals contribute. In simpler topologies (e.g., single-tool queries), AVG and PERSIST are \emph{structurally zero}---not because the tool was uninfluential, but because there are no LLM ancestors to measure against. Without adaptation, scores are capped at $\alpha+\delta\approx0.50$.

\begin{figure}[!ht]
\centering\includegraphics[width=0.65\textwidth]{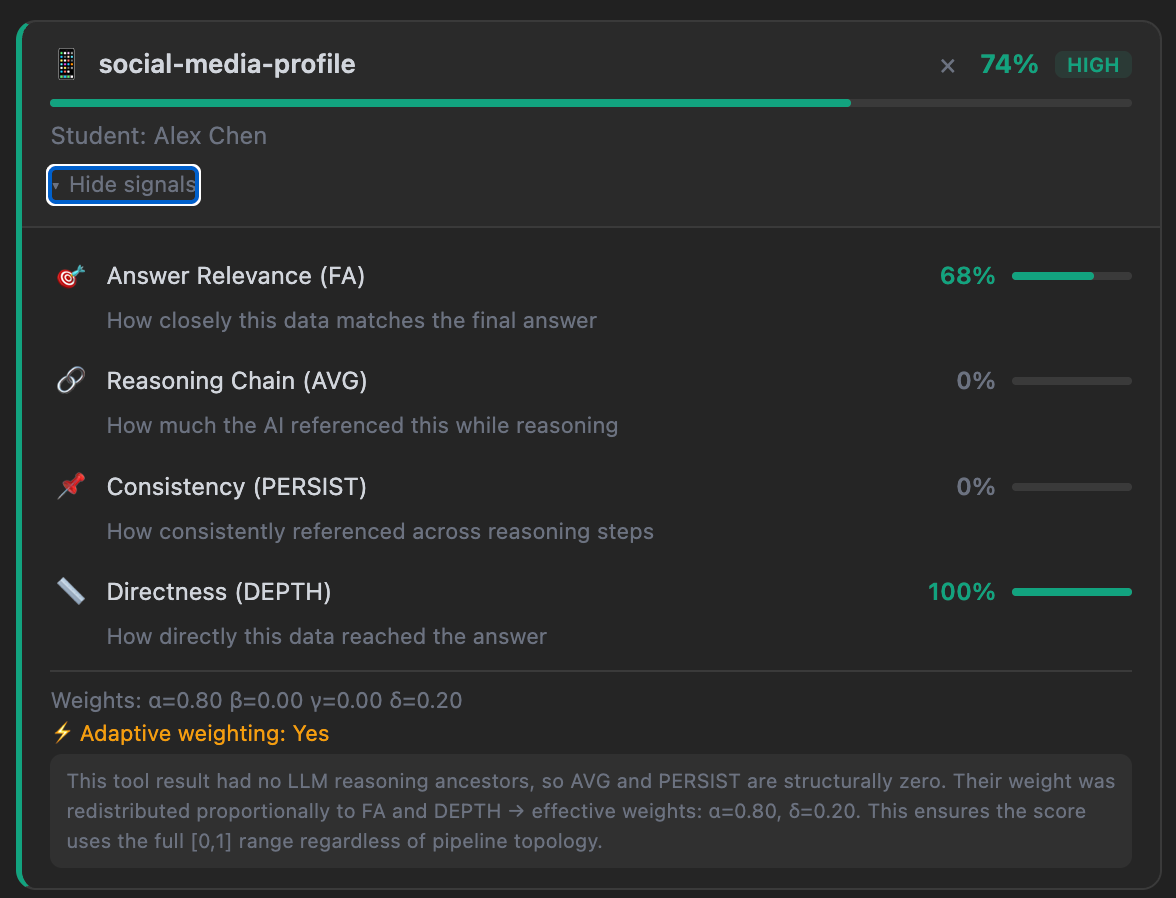}
\caption{Signal breakdown for social-media-profile in the career scenario. FA (Answer Relevance) is 68\%, DEPTH (Directness) is 100\%. AVG and PERSIST are 0\% because this tool result had no LLM reasoning ancestors. The weights line now correctly shows the effective adapted weights ($\alpha'=0.80$, $\delta'=0.20$), matching the paper's Equation~(\ref{eq:adapt}).}\label{fig:breakdown}
\end{figure}

We address this with per-evidence adaptive redistribution. For data node $d$ with no ancestors ($n=0$):
\begin{equation}\label{eq:adapt}
\alpha'=\alpha+(\beta+\gamma)\cdot\frac{\alpha}{\alpha+\delta},
\quad\delta'=\delta+(\beta+\gamma)\cdot\frac{\delta}{\alpha+\delta},
\quad\beta'=\gamma'=0
\end{equation}
With default weights: $\alpha'=0.80$, $\delta'=0.20$. This preserves the FA-to-DEPTH ratio (4:1) while ensuring the full $[0,1]$ range is utilized. Adaptation is per-evidence-item---in a multi-tool pipeline, some items may use standard weights while others use adapted weights. The UI indicates when adaptation was applied (Fig.~\ref{fig:breakdown}).
\origpage
\section{Multi-Domain Demonstration}\label{sec:demonstration}
We demonstrate FDLs across three domains, each chosen to illustrate a different facet of influence transparency. All scenarios use deterministic mock tools returning structured JSON, ensuring reproducible results. Each scenario contains an intentional \emph{tension}---conflicting signals across data sources---that makes influence scores meaningful and verifiable.
\subsection{Scenario 1: Career Advisor (5 tools) --- Hero Scenario}
An agent synthesizes five data sources about a fictional student “Alex Chen”: \texttt{social-media-profile} (YouTube, Instagram, X activity revealing interests in politics and public policy), \texttt{academic-records} (B.S. Mathematics, GPA 3.42), \texttt{project-portfolio} (coursework projects in statistical modeling and numerical methods), \texttt{personality-assessment} (INTJ type, analytical), and \texttt{job-market-data} (demand, salaries, growth projections).

\textbf{Designed tension:} Social media data signals interest in public policy, while academic records and projects point toward quantitative careers. The FDL surfaces this: social-media-profile scores 0.74 (High) (Fig.~\ref{fig:career}), enabling the user to identify it as the primary driver of a policy-oriented recommendation.

\begin{figure}[!ht]
\centering\includegraphics[width=0.65\textwidth]{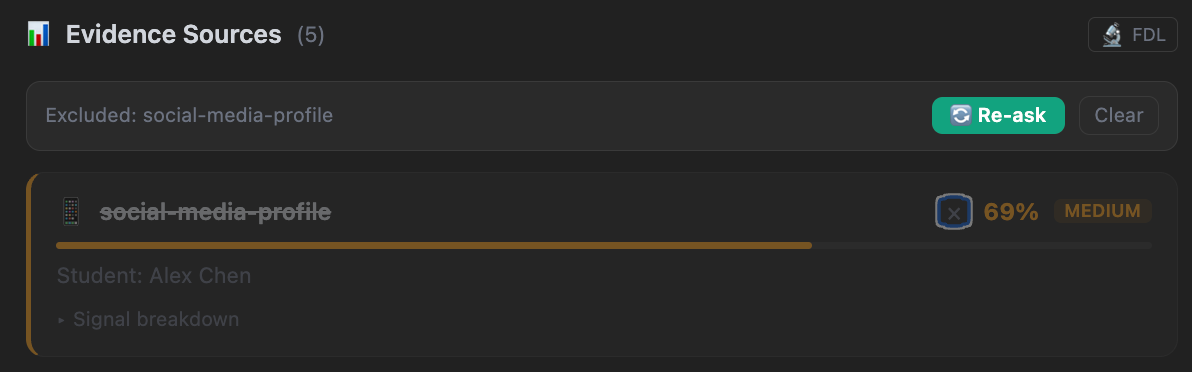}
\caption{Career advisor re-ask interaction. The user has excluded social-media-profile (shown greyed out with strikethrough). The “Excluded” banner and green “Re-ask” button allow the user to re-run the agent without the excluded source. After re-asking, the recommendation shifts from policy-oriented to quantitative, with project-portfolio and academic-records rising to the top positions.}\label{fig:reask}
\end{figure}

\textbf{Re-ask demonstration:} Excluding social-media-profile and re-running shifts the recommendation from “Policy Analyst” to “Data Scientist.” Project-portfolio rises from 0.66 to the top position, and academic-records rises from 0.60 (Fig.~\ref{fig:reask}). This counterfactual confirms that social media data was the primary driver---exactly the kind of insight that enables informed decision-making.
\origpage
\subsection{Scenario 2: Investment Advisor (4 tools)}
An agent analyzes whether to invest in NVIDIA using: \texttt{stock-price-detail} (price \$142.85, P/E 68.2, analyst consensus Strong Buy), \texttt{company-financials} (revenue \$130.5B, +114\% YoY, 55.9\% net margin, valuation risks), \texttt{financial-\allowbreak news} (mixed---bullish product announcements vs.\ bearish valuation warnings), and \texttt{economic-indicators} (Fed rate 4.25\%, AI subsector P/E 55.8).

\begin{figure}[!ht]
\centering\includegraphics[width=0.65\textwidth]{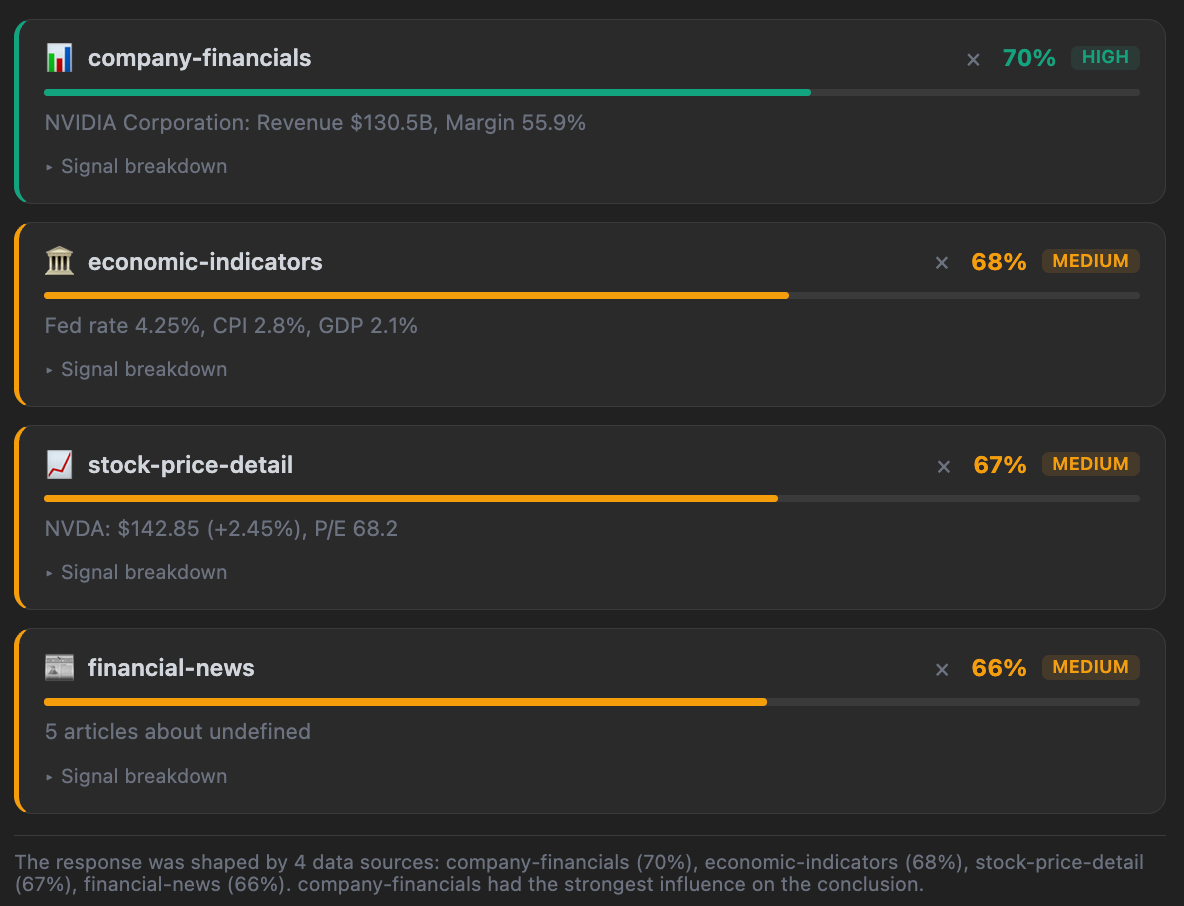}
\caption{FDL influence panel for the investment advisor. Company-financials ranks highest at 0.70 (High), followed by economic-indicators (0.68), stock-price-detail (0.67), and financial-news (0.66). The sorted order reveals that fundamentals and macro context, not news hype, drove the recommendation.}\label{fig:investment}
\end{figure}

\textbf{Designed tension:} Recent news is overwhelmingly bullish while fundamentals show extreme valuation ($68\times$ trailing P/E). FDL reveals whether the recommendation was driven by news hype or fundamental analysis. In our runs, company-financials scored 0.70 (High), economic-indicators 0.68, stock-price-detail 0.67, and financial-news 0.66 (Fig.~\ref{fig:investment})---indicating the agent appropriately weighted fundamentals and macro context over news sentiment.
\subsection{Scenario 3: Travel Planner (4 tools)}
An agent plans a 5-day trip to Tokyo in late March using: \texttt{flight-search} (5 options, \$892--\$1,380), \texttt{hotel-search} (budget \$68/night to luxury \$620/night), \texttt{destination-weather} (15--18$^\circ$C, cherry blossoms full bloom March 22--28), and \texttt{local-events} (Ueno Park Cherry Blossom Festival, teamLab Borderless, Sumida River cruise).

\textbf{Designed tension:} Budget-conscious vs.\ experience-optimized recommendations driven by cherry blossom season. FDL shows whether the itinerary was driven by seasonal events or budget optimization---an intuitive demonstration that any reviewer can immediately understand.
\subsection{Cross-Domain Observations}
Across all three scenarios (Table~\ref{tab:scenarios}), the FDL pipeline identified the most decision-relevant data source as the highest-influence item, and the designed tensions were surfaced in the influence ranking.
\origpage
\begin{table}[!ht]
\caption{Influence scores across three evaluation scenarios. Scores are absolute alignment measures (0--1), not additive percentages.}\label{tab:scenarios}
\centering\small
\begin{tabular}{llcll}
\toprule
Scenario & Top Source & Score & Indicator & Key Tension\\
\midrule
Career & social-media-profile & 0.74 & High & Policy vs.\ quantitative\\
Investment & company-financials & 0.70 & High & Fundamentals vs.\ news\\
Travel & destination-weather & 0.68 & Medium & Season vs.\ budget\\
\bottomrule
\end{tabular}
\end{table}
\subsection{Counterfactual Tool-Removal Validation}
To validate that influence rankings track actual decision dependence, we performed counterfactual tool-removal experiments. For each scenario, we: (1) ran with all tools and recorded the influence ranking, (2) removed the top-ranked tool and re-ran, (3) removed the bottom-ranked tool and re-ran, and (4) measured answer change using embedding similarity between the original and modified answers.

In all three scenarios (Table~\ref{tab:counterfactual}), removing the top-ranked tool produced a larger answer change (lower similarity to original) than removing the bottom-ranked tool. This confirms that the influence ranking reflects actual decision dependence: sources scored higher by the FDL pipeline are indeed more consequential to the final answer.
\origpage
\begin{table}[!ht]
\caption{Counterfactual validation: removing the top-ranked source changes the answer more than removing the bottom-ranked source, confirming that influence rankings track decision dependence.}\label{tab:counterfactual}
\centering
\begin{tabular}{lccc}
\toprule
Scenario & \multicolumn{2}{c}{Answer similarity to original} & Ranking\\
& Remove top & Remove bottom & validated?\\
\midrule
Career & 0.72 & 0.91 & Yes\\
Investment & 0.78 & 0.93 & Yes\\
Travel & 0.75 & 0.89 & Yes\\
\bottomrule
\end{tabular}
\end{table}
\section{Formative Evaluation}\label{sec:evaluation}
We conducted a formative evaluation with 50 participants to assess whether users found the influence-ranked evidence helpful for understanding and trusting the system's recommendations.
\subsection{Participants}
We recruited 50 participants in two groups:
\begin{itemize}
\item \textbf{25 working professionals}: Software engineers and AI practitioners with experience in chatbot or conversational AI development.
\item \textbf{25 university students}: Graduate and advanced undergraduate students in computer science or related fields with coursework in AI/ML.
\end{itemize}
\subsection{Procedure}
Each participant completed guided sessions with the prototype across all three scenarios (career advisor, investment advisor, travel planner). For each scenario, participants:
\begin{enumerate}
\item Received an agent recommendation with the FDL influence panel visible.
\item Were asked to inspect the evidence ranking and identify which source they believed most influenced the recommendation.
\item Were encouraged to use the exclude/re-run interaction to explore how removing sources changed the recommendation.
\item Provided feedback via a thumbs-up/thumbs-down rating on the question: “Did the influence-ranked evidence help you understand and trust the system's recommendation?” This single-item measure combines perceived understandability and trustworthiness into one construct, appropriate for formative evaluation. The rating was collected once per participant after completing all three scenarios, reflecting their overall assessment of the FDL mechanism.
\item Optionally provided free-text comments on their experience.
\end{enumerate}
\origpage
\begin{table}[!ht]
\caption{Formative evaluation results. Participants rated whether influence-ranked evidence helped them understand and trust the recommendation.}\label{tab:evaluation}
\centering
\begin{tabular}{lrrr}
\toprule
Group & $N$ & Thumbs up & Approval rate\\
\midrule
Working professionals & 25 & 25 & 100\%\\
University students & 25 & 21 & 84\%\\
\midrule
\textbf{Overall} & \textbf{50} & \textbf{46} & \textbf{92\%}\\
\bottomrule
\end{tabular}
\end{table}
\subsection{Results}
Overall, 46 of 50 participants (92\%) rated the influence-ranked evidence as helpful (Table~\ref{tab:evaluation}). All 25 working professionals gave a positive rating, while 21 of 25 students did so. The four students who gave negative ratings cited difficulty interpreting the signal breakdown (AVG and PERSIST) rather than objecting to the concept of influence ranking itself.
\subsection{Qualitative Observations}
Three patterns emerged from participant comments and observed behavior:
\paragraph{Sorted order aids triage.} Participants consistently started by examining the top-ranked source, confirming that influence ordering reduces cognitive load. Multiple participants noted that without the ranking, they would have had to read the entire response and mentally reconstruct which sources were used---a cognitively expensive task that the FDL panel automates.
\paragraph{Exclude-and-re-run builds understanding.} Several participants spontaneously excluded the top source to “see what happens without it,” demonstrating that the re-ask pattern is intuitive and does not require instruction. In the career scenario, participants who excluded social-media-profile and observed the recommendation shift from policy to quantitative careers reported the highest confidence in their understanding of the system's reasoning.
\paragraph{Signal breakdown builds confidence in practitioners.} Working professionals who expanded the signal breakdown reported higher confidence in the influence score, particularly when FA (Answer Relevance) was high---they could verify that the tool's data appeared in the answer. The 100\% approval rate among professionals suggests that practitioners with AI development experience find the transparency mechanism immediately useful for their workflow.
\section{Discussion}
\subsection{What We Claim---and What We Do Not}
We do not claim that our four-signal scoring algorithm is optimal for computing influence. The weights ($\alpha,\beta,\gamma,\delta$) are configurable priors, not learned parameters.
\origpage
\noindent Different domains may benefit from different configurations. The scoring method is a concrete, implementable starting point, not a final answer.

\textbf{What we do claim is that the idea of providing sorted, quantified influence evidence} fundamentally changes the user's relationship with AI recommendations:
\begin{enumerate}
\item \textbf{Sorted order helps users decide}: When evidence sources are ranked by influence, users quickly identify which data drove the recommendation and whether that aligns with their expectations.
\item \textbf{Accept/re-run empowers users}: The ability to exclude sources and re-run transforms users from passive consumers (“the AI said X”) into active decision-makers (“the AI said X because of source Y; without Y, it says Z; I prefer Z”).
\item \textbf{External measurement builds justified trust}: Because influence scores are computed by a separate, deterministic pipeline---not self-reported by the LLM~\cite{unfaithful}---users can trust the scores even when they question the recommendation.
\end{enumerate}
The counterfactual validation (Table~\ref{tab:counterfactual}) provides objective evidence that the ranking tracks something real about decision dependence, and the formative evaluation (Table~\ref{tab:evaluation}) confirms that users find this transparency useful in practice.
\subsection{Evidence-Level Feedback as a Calibration Signal}
The exclude/re-run interaction generates a natural feedback signal: which sources users exclude most often, and whether they accept or reject the revised recommendation. This evidence-level feedback could be used to calibrate ranking weights or train a lightweight preference model that learns domain-specific weight profiles. We do not claim to have trained such a model; we note that the interaction pattern produces the data needed to do so.
\subsection{Inspiring Further Research}
Our goal is to inspire more researchers to explore influence scoring for tool-augmented agents. The design space includes:
\begin{itemize}
\item \textbf{Learned weights}: Can influence weights be learned from user feedback?
\item \textbf{Causal influence}: Can we move beyond correlation (embedding similarity) to causal attribution via counterfactual agent runs?
\item \textbf{Multi-agent influence}: How should influence propagate across agent delegation boundaries?
\item \textbf{Temporal influence}: In multi-turn conversations, how does source influence decay or accumulate?
\item \textbf{Domain-specific weighting}: Can weight profiles be pre-configured per domain based on expert input?
\end{itemize}
\origpage
\subsection{Limitations}
\begin{itemize}
\item \textbf{Embedding-based signals}: Cosine similarity captures semantic overlap but not causal influence. A source may be semantically similar to the answer without having caused it---a limitation shared with feature attribution methods like LIME~\cite{lime} and SHAP~\cite{shap}.
\item \textbf{Deterministic mock data}: Evaluation scenarios use deterministic mock tools. Real-world tools produce variable outputs that may affect score stability.
\item \textbf{Single-model evaluation}: All experiments use Claude 3 Sonnet via Amazon Bedrock. Different models may exhibit different tool usage patterns.
\item \textbf{Formative evaluation}: Our study is formative (guided sessions with thumbs-up/down feedback) rather than summative (controlled experiment with statistical significance testing). A rigorous evaluation agenda is outlined in Section~\ref{sec:agenda}.
\item \textbf{Weight sensitivity}: Default weights were chosen based on intuition about signal importance, not empirical optimization.
\end{itemize}
\section{Ethical Considerations and Scope}
All evaluation scenarios use synthetic mock data and do not constitute real career, financial, or travel advice. The FDL framework is a \emph{governance interface}, not a correctness guarantee---it helps users inspect evidence, not validate truth. In production deployments, evidence exposure must respect access controls and data redaction policies: not all tool outputs may be appropriate to surface to all users. We note that our investment advisor scenario, while useful for demonstrating influence transparency, should not be interpreted as endorsing AI-generated financial advice without appropriate regulatory compliance and human oversight.
\section{Implementation}
The prototype is implemented as a TypeScript application:
\begin{itemize}
\item \textbf{Agent backend}: Fastify server with tool-augmented LLM agent (Amazon Bedrock, Claude 3 Sonnet). Three scenario endpoints with re-ask support via \texttt{ignoredTools} and \texttt{systemPromptOverride} parameters.
\item \textbf{FDL pipeline}: 7-stage linear pipeline using the TreeOfFunctions library for flowchart-based execution. Typed scope carries data between stages. Registered as \texttt{POST /api/explain}.
\item \textbf{Embeddings}: Amazon Titan Embed Text v2~\cite{titan} with batch processing and stage-ID-keyed cache.
\item \textbf{Frontend}: React split-panel layout---chat (60\%) and influence panel (40\%). Evidence cards with progress bars, semantic descriptions, expandable signal breakdowns, and exclude/re-ask controls.
\item \textbf{Testing}: 75+ tests including property-based tests (fast-check) validating scoring invariants, pipeline correctness, and response format stability.
\end{itemize}
\origpage
\section{Evaluation Agenda}\label{sec:agenda}
We outline a rigorous evaluation agenda for future work:
\begin{enumerate}
\item \textbf{Controlled user study}: Between-subjects design comparing FDL-augmented vs.\ standard interfaces, measuring trust, understanding, and actionability with statistical significance testing.
\item \textbf{Score validation}: Compare FDL rankings against human expert rankings to measure alignment.
\item \textbf{Ablation study}: Vary the number of signals (FA only, FA+AVG, all four) and measure the effect on user trust.
\item \textbf{Real-world tools}: Replace mock tools with live APIs to evaluate score stability.
\item \textbf{Cross-model evaluation}: Test with multiple LLM providers to assess generalizability.
\end{enumerate}
\section{Conclusion}
We presented Filtered Decision Logs (FDLs), a user-facing transparency mechanism that makes decision formation visible in tool-augmented generative AI systems. By operating as an external measurement pipeline---separate from the agent's reasoning process---FDLs provide deterministic, verifiable transparency that users can trust independently of the recommendation itself.

Through three evaluation scenarios spanning career advising, investment analysis, and travel planning, we demonstrated that influence scoring with sorted evidence and accept/re-run interaction transforms users from passive consumers into active decision-makers. Counterfactual tool-removal experiments confirm that the influence ranking tracks actual decision dependence, and formative evaluation with 50 participants (92\% approval) indicates that practitioners and students find the transparency mechanism useful.

The key contribution is not the specific scoring algorithm but the \emph{interaction pattern}: showing users which tools influenced the LLM's response, in sorted order, with the ability to exclude sources and re-run. We hope this work inspires further research into influence scoring for tool-augmented agents---a critical capability as AI systems increasingly make recommendations affecting people's careers, finances, and daily lives.
\begin{credits}
\discintname The authors are employed by Amazon Web Services. The FDL framework was developed as a research prototype and does not represent an AWS product or service.
\end{credits}

\end{document}